\documentclass[a4paper,11pt]{article}
\pdfoutput=1
\usepackage{jheppub}
\usepackage[T1]{fontenc}

\title{\boldmath Entanglement Entropy of $AdS_5 \times S^5$ with massless flavors at non-zero temperature}
\author[a]{Sen Hu,}
\author[a]{Guozhen Wu}
\affiliation[a]{Wu Wen-Tsun Key Lab of Mathematics of Chinese Academy of Sciences,\\School of Mathematical Sciences,\\University of Science and Technology of China, Hefei, Anhui 230026, China}
\emailAdd{shu@ustc.edu.cn}
\emailAdd{kcwoo@mail.ustc.edu.cn}

\abstract{We consider backreacted $AdS_5 \times S^5$ coupled with $N_f$ massless flavors introduced by D7-branes at non-zero temperature. The backreacted geometry is in the Veneziano limit. The temperature of this system is related to the event horizon at $r_h$. Dividing one of the spatial directions into a line segment with length $l$, we will calculate the entanglement entropy between the two subspaces. We study the behavior near the event horizon, and finally find that there exists phase transition phenomenon near the event horizon since the difference between the entanglement entropy of the connected minimal surface and the disconnected one changes sign.}

\begin{document}
\maketitle
\flushbottom

\section{Introduction}
\label{sec:1}

In a given system, entanglement entropy quantifies how closely one of the subsystems would be related to another. The entanglement entropy (EE) is defined as the von Neumann entropy of the reduced density matrix $\rho_A$ of the subsystem A where the observer can only get access to:
\begin{equation}
S_A=-tr\rho_A \log\rho_A\,.
\end{equation}
The EE is also related to the degrees of freedom in a system.

Traditionally, in order to calculate the EE, we will use the Renyi entropy and then apply the replica trick. In \cite{1,2} the authors propose a powerful method to compute the EE holographically. The EE between two subsystems A and its complement B is proportional to the boundary separates them which is known as the area law:
\begin{equation}
S_A=\frac{Area(\gamma_A)}{4G_N^{d+2}}\,.
\end{equation}
Notice that the result of the EE is divergent, and it is not invariant under a UV cut-off rescaling.

In this paper, we will follow the procedure of \cite{3} in which the authors compare the difference of EE related to connected and disconnected minimal surface respectively and find the smaller one. If the difference of EE changes sign, it means that there exists confinement/deconfinement phase transition. In addition, by comparing the difference of EE, one can find the UV cut-off independent part of the EE.

In this paper, we will consider $SU(N_c)$ $\mathcal{N} =4$ SYM coupled with $N_f$ massless flavors in the Veneziano limit at non-zero temperature. After the introduction of $N_f\gg1$ massless flavors, the smearing procedure is applied. The background metric will be calculated up to first order of $\epsilon_h$, $\epsilon_h\sim\lambda_hN_f/N_c$, where $\lambda_h$ is the 't Hooft coupling at $r_h$ which is proportional to the temperature of the system \cite{4}. By dividing the $d$ dimensional space into two complementary subspaces with a strip of length $l$, we will compute the EE between these two regions. By comparing the difference of the EE with respect to the connected and disconnected minimal surface, we find that near the event horizon the EE related to disconnected minimal surface is smaller than the connected one. When the radial coordinate $r$ gets larger, the EE of connected minimal surface is then smaller than that of the disconnected one. The changes of different shape of the minimal surface implies that there exists phase transition related to confinement/deconfinement. Notice that the EE is given by the minimization of the area action, so we will always choose the smaller EE.

To get back the phase transition in the system of $AdS$, one can consider the hard and soft wall models which introduce hard-wall and soft-wall cut-off in the IR region (see the review of \cite{5}). In this paper, we get back the phase transition
by considering the non-zero temperature of the system and find that after introducing the non-zero temperature we get back the phase transition near the event horizon.

The paper is structured as follows. In section \ref{sec:2}, we will introduce the work of \cite{3} and calculate the $AdS_5 \times S^5$ at non-zero temperature as an example. We can find that there exists phase transition phenomenon near the event horizon. In section \ref{sec:3}, we will introduce the work of \cite{4} which will be served as the background metric. We choose to study the near horizon behavior of the system and find that there exists phase transition near the event horizon since the difference of EE related to connected and disconnected minimal surface changes sign. In section \ref{sec:4}, we will make conclusion on the results and show some possible work in the future.

\section{Entanglement entropy computation}
\label{sec:2}
\subsection{Calculation of entanglement entropy}
\label{sec:2.1}

In this subsection, we will give a summarization of the work in \cite{3}. In that paper, the authors generalize the work of \cite{1,2} to large $N_c$ theories.

If a system is divided into two complementary subsystems, and an observer is located in a subsystem A and gets no access to its complement B, the EE between two subsystems is the quantity that measures the entropy the observer can get from the entanglement between region A and B. In the paper of \cite{1,2}, the authors propose a convenient method to compute the EE which is known as the area law:
\begin{equation}
S_A=\frac{Area(\gamma_A)}{4G_N^{d+2}}\,.
\end{equation}
In \cite{3}, the authors give a proposal of a more generalized method of EE computation:
\begin{equation}
S_A=\frac{1}{4G^{(10)}_{N}}\int d^8\sigma e^{-2\Phi}\sqrt{G^{(8)}_{ind}}\,.
\end{equation}
Notice that we compute the EE of two complementary regions by minimizing the area action above.

The background metric we will consider is:
\begin{equation}
{ds}^{2}_{10}=\alpha(r)(\beta(r){dr}^2+{dx}^{\mu}{dx}_{\mu})+g_{ij}{dy}^i{dy}^j\,,
\end{equation}
where $r$ is the radial coordinate ranging from $r_0$ to $\infty$, $x^\mu (\mu=0,1,\cdots,d)$ is the parameters of $\mathbb{R}^{d+1}$, $y^i (i=d+2,\cdots, 9)$ is the parameters of $8-d$ internal manifold.
The volume of the internal manifold is given by:
\begin{equation}
V_{int}=\int\prod^{8-d}_{i=1}dy^i\sqrt{\det g}\,.
\end{equation}
We will introduce a new function:
\begin{equation}
H(r)=e^{-4\Phi}V^{2}_{int}\alpha^d\,.
\end{equation}

The shape of subsystem A we will consider is the straight belt with length $l$.
The calculation of EE is similar to the original paper \cite{1,2}:
\begin{equation}
\frac{S_A}{V_{d-1}}=\frac{1}{4G^{(10)}_{N}}\int^{\frac{l}{2}}_{-\frac{l}{2}}dx\sqrt{H(r)}\sqrt{1+\beta(r)(\partial_xr)^2}\,.
\end{equation}
The possible minimal surfaces have two options, and we will choose the one with smaller EE: one is a disconnected surface which consists of two cigar-like surfaces that extend in $\mathbb{R}^{d-1}$ and separate in the remaining direction of  $\mathbb{R}^{d}$ by the distance $l$; the other is like a tube which connects the two cigar-like surfaces.

The entanglement length $l$ with respect to the connected surface is:
\begin{equation}
\label{eq:2.7}
l(\tilde{r})=2\sqrt{H(\tilde{r})}\int^{\infty}_{\tilde{r}}\frac{dr\sqrt{\beta(r)}}{\sqrt{H(r)-H(\tilde{r})}}\,,
\end{equation}
where $\tilde{r}$ is the minimal value of $r$ related the turning point of the minimal surface.
The EE of the connected entanglement surface is:
\begin{equation}
\label{eq:2.8}
\frac{S_C(\tilde{r})}{V_{d-1}}=\frac{1}{2G^{(10)}_{N}}\int^{\infty}_{\tilde{r}}dr\frac{\sqrt{\beta(r)}H(r)}{\sqrt{H(r)-H(\tilde{r})}}\,.
\end{equation}
The EE of the disconnected one is:
\begin{equation}
\label{eq:2.9}
\frac{S_D}{V_{d-1}}=\frac{1}{2G^{(10)}_{N}}\int^{\infty}_{r_0}dr\sqrt{\beta(r)H(r)}\,.
\end{equation}
Both of EE of the connected and disconnected surface are UV divergent and dependent of the UV cut-off, but the difference between them is finite and UV cut-off independent:
\begin{equation}
\label{eq:2.10}
\frac{S_C-S_D}{V_{d-1}}=\frac{1}{2G^{(10)}_{N}}\int^{\infty}_{\tilde{r}}dr\sqrt{\beta(r)H(r)}\left(\frac{1}{\sqrt{1-\frac{H(\tilde{r})}{H(r)}}}\right)-\int^{\tilde{r}}_{r_0}dr\sqrt{\beta(r)H(r)}\,.
\end{equation}
Notice that the result of (\ref{eq:2.10}) may change sign which indicates a phase transition at the critical point. That means the preferred shape of entanglement surface changes between disconnected and connected. We will focus on the quantity of the difference of EE between the connected and disconnected entanglement surface in the remaining of this paper.

\subsection{Entanglement entropy of $AdS_5\times S^5$ at non-zero temperature}
\label{sec:2.2}

In this subsection we will compute the entanglement length and EE of $AdS_5\times S^5$ at non-zero temperature.

The metric of $AdS_5\times S^5$ at non-zero temperature is:
\begin{equation}
{ds}^2=\frac{R^2}{r^2}\frac{{dr}^2}{f(r)}+\frac{r^2}{R^2}(-f(r){dt}^2+{\vec{dx}}^{2}_{3})+R^2{d\Omega}^{2}_{5}\,,
\end{equation}
where $f(r)=1-r^{4}_{h}/r^4$. $r_h$ is set as the horizon radius and is proportional to the Hawking temperature. Sending $r_h$ to zero, we recover the pure $AdS_5\times S^5$ at zero temperature.

The relevant functions to be used later are listed as follows:
\begin{equation}
\label{eq:2.12}
\beta(U)=\frac{R^4}{r^4}\frac{1}{f(r)},\quad H(r)=\left(\frac{8\pi^2}{3}\right)^2R^4r^6\,,
\end{equation}
where $r$ ranges from $r_h$ to $\infty$.

\begin{figure}[bhtp]
\centering
\includegraphics[width=.45\textwidth,clip]{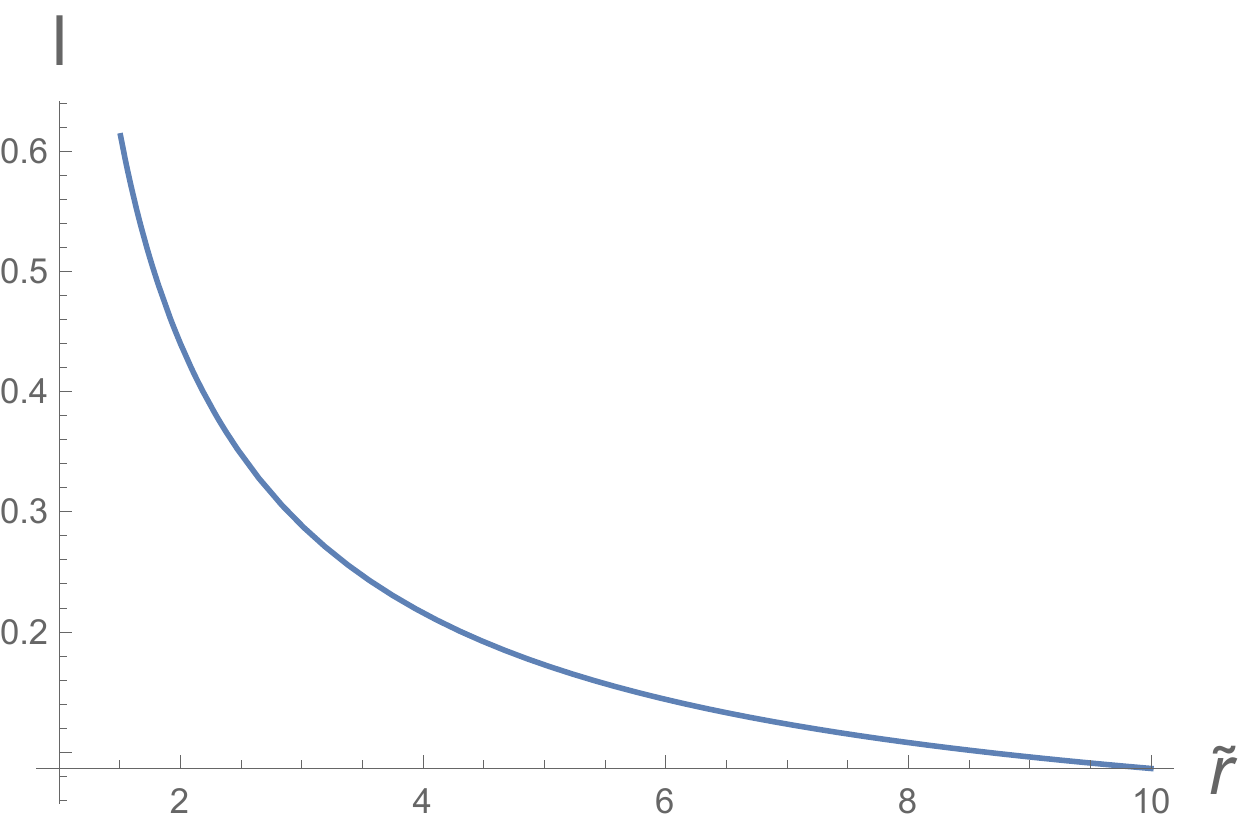}
\caption{\label{fig:1} The entanglement length as the function of $\tilde{r}$}
\end{figure}

The entanglement length $l$ is:
\begin{equation}
\label{eq:2.13}
l(\tilde{r})=2R^2{\tilde{r}}^{3}\int^{\infty}_{\tilde{r}}\frac{1}{\sqrt{(r^4-r^{4}_{h})(r^6-{\tilde{r}}^{6})}}\,,
\end{equation}
where $\tilde{r}$ is the minimal value of $r$ related to the connected surface and we have the following relation: $0<r_h\leq {\tilde{r}}\leq r<+\infty$. Notice that $l(r)$ is divergent at $r=r_h$ and is a monotonically decreasing function which goes to zero at infinity. This can be seen easily from (\ref{eq:2.13}) that the leading order of $l(\tilde{r})$ is $1/\tilde{r}$. Figure \ref{fig:1} shows the plot of $l$ as a function of $\tilde{r}$. Notice that we have set the scale of horizon $r_h$ and radius $R$ to 1.
\\
The EE of connected minimal surface is:
\begin{equation}
\begin{split}
S_C(r)&\sim\int^{+\infty}_{\tilde{r}}dr\frac{\sqrt{\beta(r)}H(r)}{\sqrt{H(r)-H({\tilde{r}})}}\\
&=R^4(\frac{8{\pi}^2}{3})\int^{+\infty}_{\tilde{r}}dr\frac{r^6}{\sqrt{(r^6-{\tilde{r}}^{6})(r^4-{{r}}^{4}_{h})}}\,.
\end{split}
\end{equation}
The EE of disconnected minimal surface is:
\begin{equation}
\begin{split}
S_D&\sim\int^{+\infty}_{r_h}dr\sqrt{\beta(r)H(r)}\\
&=R^4(\frac{8{\pi}^2}{3})\int^{+\infty}_{r_h}dr\sqrt{\frac{r^6}{r^4-{{r}}^{4}_{h}}}\,.
\end{split}
\end{equation}
The difference of them is:
\begin{equation}
\begin{split}
\label{eq:2.16}
S_C-S_D&\sim\int^{+\infty}_{\tilde{r}}dr\frac{r^6}{\sqrt{(r^6-{\tilde{r}}^{6})(r^4-{{r}}^{4}_{h})}}-\int^{+\infty}_{r_h}dr\sqrt{\frac{r^6}{r^4-{{r}}^{4}_{h}}}\\
&=\int^{+\infty}_{\tilde{r}}dr\frac{r^3}{\sqrt{r^4-r^{4}_{h}}}\frac{r^3}{\sqrt{r^6-\tilde{r}^{6}}}-\int^{+\infty}_{r_h}dr\frac{r^3}{\sqrt{r^4-r^{4}_{h}}}\,.
\end{split}
\end{equation}
Notice that when $r\rightarrow+\infty$, $\frac{r^3}{\sqrt{r^6-{\tilde{r}}^{6}}}\rightarrow1$.
When $r\rightarrow+\infty$, although the first term and second term in (\ref{eq:2.16}) both diverge, the divergent terms are the same and thus cancel. This coincides with the result of \cite{3} that the divergent part cancels at the boundary, hence we get a result independent of the UV cut-off.
To be more precise,
\begin{equation*}
\int^{+\infty}_{r_h}dr\sqrt{\frac{r^6}{r^4-{r}^{4}_{h}}}=\frac{\sqrt{r^{4}-r^{4}_{h}}}{2}{\Bigg|}^{+\infty}_{r_h}
\end{equation*}
is the shared divergent term.

In the following analysis, we will take the scale $r_h$ related to the temperature equal to 1 to simplify our computation.

The next step we will take is to compare the difference of $S_C$ and $S_D$.

Let us consider the case when $\tilde{r}\rightarrow1(=r_h)$ first. This is the near horizon case. (\ref{eq:2.16}) becomes:
\begin{equation}
\label{eq:2.17}
\int^{+\infty}_{1}dr\frac{r^3}{\sqrt{r^4-1}}\frac{r^3}{\sqrt{r^6-1}}-\int^{+\infty}_{1}dr\frac{r^3}{\sqrt{r^4-1}}\,.
\end{equation}
Since $\frac{r^3}{\sqrt{r^6-1}}>1$, the result of (\ref{eq:2.17}) is larger than zero which means that the EE of connected minimal surface is larger than that of the disconnected one at near horizon. Notice that we always choose the minimal surface with respect to a smaller EE, hence in this case the disconnected minimal surface is preferred.

The second case is when $\tilde{r}\rightarrow+\infty$. This is the near boundary case. (\ref{eq:2.16}) becomes:
\begin{equation}
\label{eq:2.18}
\int^{+\infty}_{\tilde{r}}dr\frac{r^3}{\sqrt{r^4-1}}(\frac{r^3}{\sqrt{r^6-{\tilde{r}}^6}}-1)-\int^{\tilde{r}}_{1}dr\frac{r^3}{\sqrt{r^4-1}}\,.
\end{equation}
The first term in (\ref{eq:2.18}) becomes zero, while the second term is divergent. The result of (\ref{eq:2.18}) is smaller than zero which means that the EE of connected minimal surface is smaller than that of the disconnected one, hence the connected minimal surface is preferred.

From the analysis above, we can see that as $\tilde{r}$ going from $r_h$ to $\infty$, the difference of EE of connected and disconnected minimal surface changes sign. This implies that a phase transition occurs in the bulk of black $AdS_5 \times S^5$. When $r_h$ the minimal value of $r$ is close to the horizon, $S_C-S_D>0$, $S_D$ is smaller than $S_D$ and thus preferred. In this case, $r_h\rightarrow1$ is correspondent to large $l$. When $r_h$ is closed to boundary (Actually, the phase transition occurs at near horizon, far away from the boundary. We consider the near boundary case just for convenience), $S_C-S_D>0$, $S_C$ is smaller than $S_D$ and thus preferred. In this case, $r_h\rightarrow+\infty$ is correspondent to small $l$.

In pure $AdS_5 \times S^5$, $S_C$ is always smaller than $S_D$ and thus preferred, thus we find no phase transition there \cite{5}. As we introduce the temperature into our system, the difference of $S_C$ and $S_D$ changes sign in the bulk space and we get back our phase transition. 

\section{$AdS_5 \times S^5$ with massless flavors at non-zero temperature}
\label{sec:3}
\subsection{$AdS_5 \times S^5$ with massless flavors at non-zero temperature}
\label{sec:3.1}

In this subsection, we will give a review of the work in \cite{4} (see also in \cite{6}).

Let us start with $AdS_5 \times S^5$ which is dual to $\mathcal{N}=4$ SYM. The $N_c$ D3 branes are placed at the tip of a Calabi-Yau cone over a Sasaki-Einstein manifold $S^5$. Under $U(1)$ fibration, the K\"{a}hler-Einstein
base of $S^5$ is $CP^2$, so the metric of this 5d Sasaki-Einstein manifold can be written as:
\begin{equation}
{ds}^2_{X_5}={ds}^2_{KE}+(A_{KE})^2\,,
\end{equation}
where $A_{KE}$  is the connection one form. The metric of $CP^2$ reads
\begin{equation}
{ds}^2_{{CP}^2}=\frac{1}{4}{d\chi}^2+\frac{1}{4}{\cos}^2{\frac{\chi}{2}}({d\theta}^2+{\sin}^2\theta{d\varphi}^2)+\frac{1}{4}{\cos}^2{\frac{\chi}{2}}{\sin}^2{\frac{\chi}{2}}(d\psi+{\cos\theta}d\varphi)^2\,,
\end{equation}
\begin{equation*}
A_{{CP}^2}=\frac{1}{2}{\cos}^2\frac{\chi}{2}(d\psi+{\cos\theta}d\varphi)\,,
\end{equation*}
\begin{equation*}
0\leq\chi,\theta\leq\pi,0\leq\varphi,\tau<2\pi,0\leq\psi<4\pi\,.
\end{equation*}

D7 branes are introduced as matter. The D7 branes locate along the radial direction wrapping an $S^3$ inside $S^5$ and spacetime filling in the UV region. Since we will consider the Veneziano limit, which means $N_f, N_c \rightarrow \infty$, $N_f/N_c$ is fixed. The D7 brans are smeared over the transverse space. This smearing procedure is necessary since we need to avoid the difficulty of the $\delta$ function when computing the integral of DBI action \cite{7}. After the introduction of $N_f$ D7 branes, the system preserves $\mathcal{N}=1$ SYM at zero temperature.

The DBI action is:
\begin{equation}
S=S_ {\uppercase\expandafter{\romannumeral2}B}+S_{fl}\,.
\end{equation}
The action of type ${\uppercase\expandafter{\romannumeral2}B}$ supergravity is:
\begin{equation*}
S_ {\uppercase\expandafter{\romannumeral2}B}=\frac{1}{2\kappa^{2}_{10}}\int{d^{10} x}\sqrt{-g_{10}}[R-\frac{1}{2}\partial_M\Phi\partial^M\Phi-\frac{1}{2}e^{2\Phi}F^{2}_{(1)}-\frac{1}{2}\frac{1}{5!}F^{2}_{(5)}]\,,
\end{equation*}
The action of D7 flavor branes is:
\begin{equation*}
S_{fl}=-T_7\sum_{N_f}\left(\int{d^8x}e^\Phi\sqrt{-g_8}-\int C_8\right)\,,
\end{equation*}
with the gravitational constant of
\begin{equation*}
\frac{1}{\kappa_{10}^{2}}=\frac{T_7}{g_s}=\frac{1}{(2\pi)^2g_{s}^{2}\alpha^{\prime 4}}\,.
\end{equation*}

The metric ansatz of the background is:
\begin{equation}
{ds}^{2}_{10}=h^{-\frac{1}{2}}(-b {dt}^2+{d\vec{x}}^{2}_{3})+h^{\frac{1}{2}}[bS^8F^2{d\sigma}^2+S^2{ds}^{2}_{{CP}^2}+F^2(d\tau+A^{2}_{{CP}^2})]\,.
\end{equation}
When $b=1$, we recover the zero temperature result. Notice that after the smearing procedure, the functions $h$, $b$, $S$, $F$ are only dependent on the radial coordinate $\sigma$.

The equations of motion are:
\begin{equation}
\begin{split}
\partial^{2}_{\sigma}(\log b)=&0\,,
\\
\partial^{2}_{\sigma}(\log h)=&-Q^{2}_{c}\frac{b}{h^2}\,,
\\
\partial^{2}_{\sigma}(\log S)=&-2bF^4S^4+6bF^2S^6-Q_fe^{\Phi}bF^2S^6\,,
\\
\partial^{2}_{\sigma}(\log F)=&4bF^4S^4-\frac{Q^{2}_{f}}{2}e^{2\Phi}bS^8\,,
\\
\partial^{2}_{\sigma}\Phi=&Q^{2}_{f}e^{2\Phi}bS^8+4Q_fbe^{\Phi}S^6F^2\,.
\end{split}
\end{equation}
When $b\neq1$ which is dual to non-zero temperature, the functions of $b$ and $h$ can be solved first with an integration constant $r_h$:
\begin{equation}
\begin{split}
b&=e^{4r^{4}_{h}\sigma}\,,
\\
h&=\frac{Q_c}{4r^{4}_{h}}(1-e^{4r^{4}_{h}\sigma})\,,
\end{split}
\end{equation}
where $\sigma$ ranges from $-\infty$ to 0.
The next step is to define a new variable $r$ by an implying coordinate transformation:
\begin{equation}
\begin{split}
e^{4r^{4}_{h}\sigma}&=1-\frac{r^{4}_{h}}{r^4}=b\,,
\\
h&=\frac{Q_c}{4r^4}=\frac{R^4}{r^4}\,,
\end{split}
\end{equation}
where $r$ ranges from $r_h$ to $+\infty$.

The metric of $AdS_5 \times S^5$ with massless flavors at non-zero temperature is:
\begin{equation}
\label{eq:3.8}
{ds}^{2}_{10}=-\frac{r^2}{R^2}(1-\frac{r^{4}_{h}}{r^4}){dt}^2+\frac{r^2}{R^2}d{\vec{x}}^{2}_{3}+\frac{R^2{\tilde{S}}^8{\tilde{F}}^2}{r^2}\frac{dr^2}{(1-\frac{r^{4}_{h}}{r^4})}+R^2{\tilde{S}}^2ds^{2}_{KE}+R^2{\tilde{F}}^2(d\tau+A_{KE})\,,
\end{equation}
where
\begin{equation}
\begin{split}
\tilde{S}&=\frac{S}{r}\,,
\\
\tilde{F}&=\frac{F}{r}\,.
\end{split}
\end{equation}
We will expand the solutions of the functions of $\tilde{F}$ and $\tilde{S}$ up to first order in $\epsilon_{\ast}$ which is a expansion parameter whose meaning is given in \cite{4}:
\begin{equation}
\begin{split}
\label{eq:3.10}
\tilde{F}&=1-\frac{\epsilon_{\ast}}{24}(1+\frac{2r^4-r^{4}_{h}}{6r^{4}_{\ast}-3r^{4}_{h}})+O(\epsilon^{2}_{\ast})\,,
\\
\tilde{S}&=1+\frac{\epsilon_{\ast}}{24}(1+\frac{2r^4-r^{4}_{h}}{6r^{4}_{\ast}-3r^{4}_{h}})+O(\epsilon^{2}_{\ast})\,.
\end{split}
\end{equation}
Notice that when $\epsilon_{\ast}=0$, we recover the flavorless case.

\subsection{Entanglement entropy computation}
\label{sec:3.2}

We will introduce anther expansion parameter $\epsilon_h$ related to the temperature:
\begin{equation}
\epsilon_{h}=\frac{Vol(X_3)}{16\pi Vol(X_5)}\frac{N_f}{N_c}\lambda_{h}=\frac{\lambda_h}{8{\pi}^2}\frac{N_f}{N_c}\,,
\end{equation}
where $\lambda_h$ is the 't Hooft coupling at the horizon $r_h$.
Since $\epsilon_{\ast}=\frac{\lambda_{\ast}}{8{\pi}^2}\frac{N_f}{N_c}$, we have \begin{equation}
\epsilon_h=\epsilon_{\ast}\frac{e^{\Phi_h}}{e^{\Phi_{\ast}}}=\epsilon_{\ast}+O({\epsilon}^{2}_{\ast})\,.
\end{equation}
The solutions in (\ref{eq:3.10}) are also dependent on an arbitrary UV cut-off scale $r_{\ast}$. The expression of $\tilde{F}$ and $\tilde{S}$ will be used only when $r<r_{\ast}$. $r_{\ast}$ is related to the UV Landau pole. To decouple the UV Landau pole, we need to take $r_{\ast}\rightarrow+\infty$. We have the relations of scale as follows:
\begin{equation}
\begin{split}
0<r_h\leq r<r_{\ast}<+\infty\,,
\qquad
r_h\ll r_{\ast}\,.
\end{split}
\end{equation}

We will compute the entanglement and EE of the flavored $AdS_5\times S^5$ with massless flavors with the shape of a strip in this section. From (\ref{eq:3.8}), we get:
\begin{equation}
\label{eq:3.14}
\beta(r)=hF^2S^8\frac{1}{r^{10}}\frac{1}{f(r)},\quad H(r)=hF^2S^8\,,
\end{equation}
where $f(r)=1-r^{4}_{h}/r^4$.
$F$ and $S$ satisfy:
\begin{equation}
\begin{split}
\label{eq:3.15}
F&=r(1-\frac{\epsilon_{h}}{24}(1+\frac{2r^4-r^{4}_{h}}{6r^{4}_{\ast}-3r^{4}_{h}})+O(\epsilon^{2}_{h}))\,,
\\
S&=r(1+\frac{\epsilon_{h}}{24}(1+\frac{2r^4-r^{4}_{h}}{6r^{4}_{\ast}-3r^{4}_{h}})+O(\epsilon^{2}_{h}))\,.
\end{split}
\end{equation}
Notice that we have changed the expansion parameter $\epsilon_{\ast}$ to $\epsilon_h$ which is related to the temperature.

Let us consider the UV boundary case first. When $r\rightarrow+\infty$, since $r<r_{\ast}$ and $r_{\ast}\gg r_h$, we have $r\sim r_{\ast}$ and thus:
\begin{equation}
\frac{2r^4-r^{4}_{h}}{6r^{4}_{\ast}-3r^{4}_{h}}\rightarrow1/3\,.
\end{equation}
From (\ref{eq:3.15}), we have $F\sim r$ and $S\sim r$. Using (\ref{eq:3.14}) and compared with (\ref{eq:2.12}), we have $\beta(r)\sim\frac{1}{r^4-r^{4}_{h}}$ and $H(r)\sim r^6$. Hence $l(\tilde{r})$, $S_C(\tilde{r})$ and $S_D$ have the similar behavior as the flavorless case.

The next case we will consider is the near horizon case, where the phase transition phenomenon is expected. Notice that the extra scale $r_{\ast}$ is related to the UV scale. To decouple the UV pole, we need to take $r_{\ast}\rightarrow+\infty$. Since $r_{\ast}$ is regarded as the UV cut-off and we want to focus on the IR behavior of the system, $r\sim r_h$, the UV cut-off $r_{\ast}$ can be thought of as the UV completion of the system. It can be discarded since the correction from the UV has mild effect on the IR behavior in the spirit of effective theory. Hence the terms with $r^{4}_{h}/r^{4}_{\ast}$ in (\ref{eq:3.15}) are neglected when we concentrate on the IR behavior. Similar discussions can be found in \cite{6} (see also in \cite{4}).

The equations in (\ref{eq:3.15}) become (in the IR region):
\begin{equation}
\begin{split}
F=r(1-\frac{\epsilon_{h}}{24}+O(\epsilon^{2}_{h}))\,,
\qquad
S=r(1+\frac{\epsilon_{h}}{24}+O(\epsilon^{2}_{h}))\,.
\end{split}
\end{equation}
\begin{figure}[bhtp]
\centering
\includegraphics[width=.45\textwidth,clip]{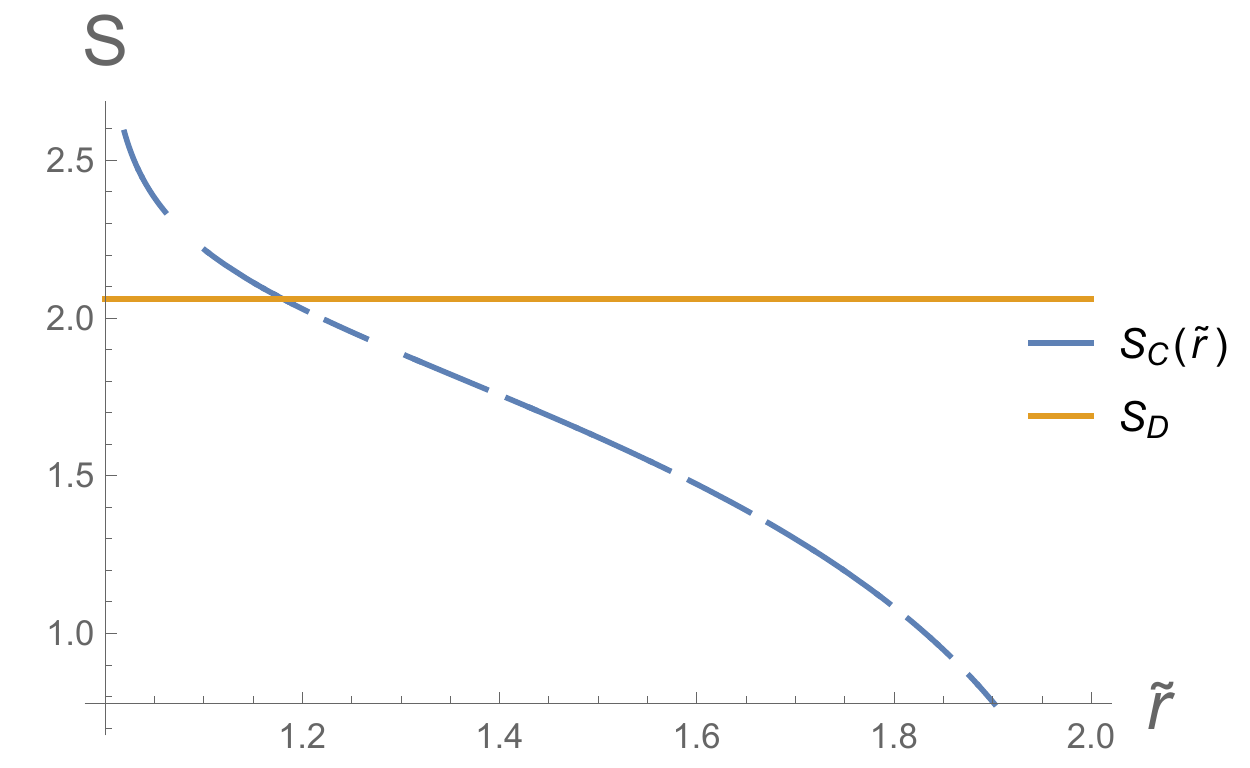}
\caption{\label{fig:2} The entanglement entropy as the function of $\tilde{r}$}
\end{figure}\\
Figure \ref{fig:2} shows the plot of $S_C$ and $S_D$ in the near horizon region. To be compared to the result of \cite{4}, we have set the value of $\epsilon_h$ to $1/4$ and $r_h$ to 1. Notice that we have the similar result as in section \ref{sec:2.2}. The EE of connected minimal surface is larger than the EE of the disconnected one when $r\rightarrow1(=r_h)$. When $r$ grows, the EE of connected surface is smaller than the disconnected one. The phase transition occurs near horizon as expected.

\section{Conclusion}
\label{sec:4}

In this paper, we have studied the entanglement entropy of $AdS_5 \times S^5$ with massless flavors in the Veneziano limit at non-zero temperature.  Dividing the $d$ dimensional space into two complementary subspaces with a strip of length $l$, we calculate the EE between them with the method of \cite{3}. We focus on the value of $S_C-S_D$ and find that there exists phase transition at near horizon $r_h$ which is related to the temperature. 
In this paper, we bring back the phase transition with the introduction of non-zero temperature.

In section \ref{sec:2}, we calculate the entanglement length and EE of $AdS_5 \times S^5$ at finite temperature. The entanglement length is monotonically decreasing and divergent at the scale of horizon $r_h$. The EE of connected minimal surface is larger than that of the disconnected one at the horizon, while the EE of connected surface is smaller than the disconnected one at the boundary, hence phase transition occurs in the bulk space which means the different choice of the shape minimal surface in the spirit of the minimization of the area action. In section \ref{sec:3}, we calculate the entanglement length and EE of $AdS_5 \times S^5$ coupled with $N_f$ massless flavors at finite temperature, finding that it shares the similar behavior with the flavorless case. The phase transition happens near the horizon. When $r_h=\tilde{r}$, $S_C>S_D$, hence the disconnected minimal surface is the preferred shape. When $r_h$ grows, the phase transition happens near the horizon, the shape of minimal surface is connected from then on.

Let us close this section by discussion about some possible future work. In  Klebanov-Strassler model, there exists phase transition in the bulk with and without flavors. In \cite{8}, the authors calculate the critical exponents using the method of Wilson loop. In \cite{9}, the authors calculate the critical exponents with the method of EE, and find that they belong to a different universality class compared to that with the method of Wilson loop. It would be interesting to find out the underlying meaning of that difference.

\acknowledgments

This work is supported by the Wu Wen-Tsun Key Laboratory of Mathematics at USTC of Chinese Academy of Sciences. The work is partially supported by NSFC under the contract nunber of 11571336.

\end{document}